# Bayesian Reconstruction of Magnetic Resonance Images using Gaussian Processes

Yihong Xu[1], Chad W. Farris[3], and Stephan W. Anderson[3], Xin Zhang[2,4,5,6], Keith A. Brown[1,2,6*]

[1]Department of Physics, Boston University, Boston, MA, USA, 02215, USA

[2]Division of Materials Science & Engineering, Boston University, Boston, MA, 02215, USA

[3]Department of Radiology, Boston Medical Center and Boston University Chobanian & Avedisian School of Medicine, Boston, MA, 02118 USA

[4]Division of Electrical & Computer Engineering, Boston University, Boston, MA, 02215, USA

[5]Division of Biomedical Engineering, Boston University, Boston, MA, 02215, USA

[6]Division of Materials Science & Engineering, Boston University, Boston, MA, 02215, USA

*Corresponding author: Prof. Keith A. Brown Email: *brownka@bu.edu*

**ORCID:**

Yihong Xu: 0000-0003-0542-4075

Chad W. Farris: 0000-0002-1133-3834

Xin Zhang: 0000-0002-4413-5084

Keith A. Brown: 0000-0002-2379-2018

**Abstract**

A central goal of modern magnetic resonance imaging (MRI) is to reduce the time required to produce high-quality images. Efforts have included hardware and software innovations such as parallel imaging, compressed sensing, and deep learning-based reconstruction. Here, we propose and demonstrate a Bayesian method to build statistical libraries of magnetic resonance (MR) images in *k*-space and use these libraries to identify optimal subsampling paths and reconstruction processes. Specifically, we compute a multivariate normal distribution based upon Gaussian processes using a publicly available library of T1-weighted images of healthy brains. We combine this library with physics-informed envelope functions to only retain meaningful correlations in *k*-space. This covariance function is then used to select a series of ring-shaped subsampling paths using Bayesian optimization such that they optimally explore space while remaining practically realizable in commercial MRI systems. Combining optimized subsampling paths found for a range of images, we compute a generalized sampling path that, when used for novel images, produces superlative structural similarity and error in comparison to previously reported reconstruction processes (i.e. 96.3% structural similarity and <0.003 normalized mean squared error from sampling only 12.5% of the *k*-space data). Finally, we use this reconstruction process on pathological data without retraining to show that reconstructed images are clinically useful for stroke identification.

# 1. Introduction

Magnetic resonance imaging (MRI) has been widely used as a non-invasive technique to image pathological or anatomical details of the internal body. A major focus in the community has been to develop approaches that reduce the time needed to obtain high quality magnetic resonance (MR) images. However, the rate at which information is obtained is limited by the instrument. One approach to circumvent this has been to parallelize the data collection process through parallel imaging (PI) in which multiple separate phase-encoded receiver channels collect parts of the image that are subsequently reconstructed to form a complete image [1-3]. Here, the maximum acceleration in imaging speed is bounded by the number of coils, which is limited in practice by physical constraints. In contrast with using multiple sensors to collect data faster, it is also possible to collect less data and then employ algorithms to reconstruct or infer a complete image. For example, the compressed sensing (CS) approach leverages the fact that typical MR images are sparse in the wavelet domain, however capitalizing on this fact is complicated by collecting in *k*-space. Thus, CS iteratively applies sparsifying transformations and denoising algorithms to obtain high quality images [4, 5]. However, the subspace of the acquired image in *k*-space has the restriction that it has to be incoherent and not produce structural aliasing [6], so these subsample patterns need to be specially designed. Besides commonly used quasi-random sampling, various non-Cartesian radial, circular, and spiral patterns have been experimentally adapted into CS frameworks [7-10]. Recently, PI and CS have been combined to reduce reconstruction artifacts and obtain higher imaging acceleration compared to either method used alone [11-13].

Fueled by the emergence of deep learning (DL) methods that have played a tremendous role in image recognition and computer vision, much work has focused on applying machine learning to further accelerate MRI imaging. For example, DL has been inserted into PI and CS algorithms to facilitate MR image reconstruction [14]. One approach to using DL involves training a neural network to predict full real-space images using subsampled *k*-space data, which has been accomplished using convolutional neural networks [15, 16], recurrent neural networks [17, 18], and parallel self-supervised networks [19]. More recently, advanced DL methods such as generative adversarial networks [20] and variational autoencoders [21] have been used to learn latent features in the data that aid in reconstructing full images from partial *k*-space data. In addition to methods that take fixed subsamples as inputs, there have also been active learning methods in which the *k*-space trajectory is dynamically chosen as the image is collected [22-24]. An important output of such work has been generalized *k*-space paths that are predicted to function well for a variety of challenging clinical applications.

While DL methods can offer promising image reconstruction performance and capacity to generalize, there exist other statistical strategies to efficiently encode information. Notably, Bayesian optimization (BO) is a powerful way of navigating complex spaces to optimally achieve a user-defined goal. This process is based on Bayes' theorem and involves iteratively building a statistical representation of a function and using this model to select additional data to collect [25]. Due to its efficient use of data, BO has been applied to accelerate a number of research areas in which measurements are expensive including material characterization and design [26-28], protein and molecular engineering [29, 30], and biomedical imaging [31-33]. Prior work has also sought to use BO to accelerate MRI. For instance, one study focused on the question of which *k*-space sampling path to adopt and used Bayesian inference to select the sampling path that was likely to produce the best final reconstruction quality [33]. However, this process neither used Bayesian methods for the reconstruction process itself nor leveraged databases of relevant MR images to provide prior knowledge in a Bayesian framework. This raises the question of whether prior knowledge could be effectively used in a Bayesian framework to both select a subsampling path and effectively perform reconstruction.

Here, we propose a method to accelerate MRI using a Bayesian framework to both select a *k*-space subsampling path and then reconstruct the subsequent subsampled *k*-space data (Figure 1). In particular, this method conditions a Gaussian process (GP) to capture the mean and covariances present in MRI data and uses the BO framework to select circular *k*-space subsampling paths that minimize the uncertainty in the reconstructed image. Images are then reconstructed using Bayesian inference. Inspired by advances in stationary kernel design [34, 35] and the physics-based Hermitian symmetry of *k*-space, we explore a series of kernel functions to modify the empirical covariance matrix and find that the use of a double-envelope function that captures the Hermitian symmetry inherent to *k*-space data allows us to obtain structural similarity indices of 0.963 with less than 12.5% of the total image. In addition, we explore the effect of training library size and find that libraries with as few as ~400 images can provide high quality reconstruction, but that larger libraries drastically reduce computational time. Strikingly, we found that this model was able to provide diagnostically useful information for pathological MR images with no additional training despite these images being generated by different MRI sequences and featuring different pathologies. Collectively, this process represents a scalable and low-bias method for reconstructing subsampled *k*-space data while leveraging symmetries present in MRI. *The code from this work is available at https://github.com/yihonglilyxu/KspaceMRIBO.*

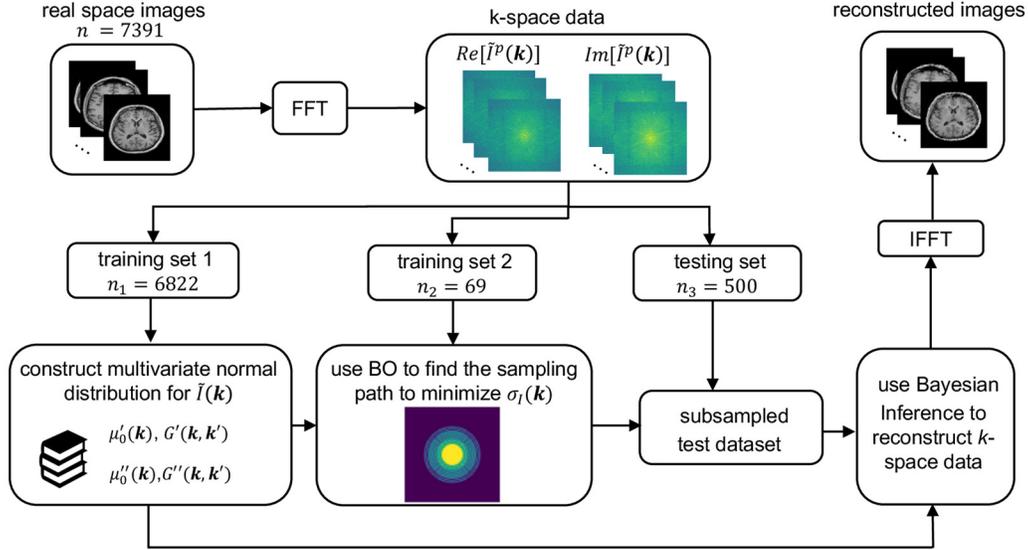

**Fig.1.** Schematic of the proposed image reconstruction process in which real space images of brains are converted to $k$-space data using a fast Fourier transform (FFT). The original set of $n$ images is broken into three subsets, subset 1 includes $n_1$ images for computing a multivariate normal distribution for the $k$-space intensity $\tilde{I}$, subset 2 includes $n_2$ images for identifying an optimal subsampling path using Bayesian optimization (BO) to minimize $k$-space error $\sigma_I$, and subset 3 includes $n_3$ images for evaluating the subsampling and reconstruction process. Final real-space images are produced using an inverse fast Fourier transform (IFFT).

## 2. Methods

This work comprises the development of a BO-based approach for optimally subsampling and reconstructing MR images. Underscoring this work is the hypothesis that there exist consistent statistical correlations between different regions in $k$-space that can be leveraged in both selecting regions for subsampling and reconstructing subsampled data. As such, this process begins by using a reference library of training images to compute a multivariate normal distribution in $k$-space. With this distribution in hand, optimized sampling paths may be identified using Bayesian optimization. Finally, the subsampled $k$-space data may be reconstructed using Bayesian inference. In order to independently evaluate each of these steps, a library of $n$ total images was separated into training set 1 with $n_1$ images to build the multivariate normal distribution, training set 2 with $n_2$ images to identify an optimized sampling path, and a testing set with $n_3$ images to evaluate reconstruction performance.

## A. Construction of Multivariate Normal Distributions

In order to begin constructing a multivariate normal distribution for the Gaussian process, we include $n_1$ real-space images ($\mathbb{R}^2$) that are each $N{\times}N$ pixels in size. For each image $p$, we use a fast Fourier transform (FFT) to convert the real-space image $I^p(\boldsymbol{x})$ into a complex $k$-space dataset $\tilde{I}^p(\boldsymbol{k})$ ($\mathbb{C}^2$), where $\boldsymbol{k}$ is the two-dimensional location in $k$-space and $\boldsymbol{x}$ is the two-dimensional location real space. Note that $\tilde{I}^p(\boldsymbol{k})$ is defined at $N{\times}N$ discrete points in $k$-space. One challenge in dealing with this data is that the magnitude of $\tilde{I}$ varies dramatically in a given image with large values near $\boldsymbol{k} = 0$ that decrease exponentially away from the origin. With this in mind, it is useful to normalize the data using the average intensity given by

$$\langle \tilde{I}(\boldsymbol{k}) \rangle_{n_1} = \sum_{p=1}^{n_1} |\tilde{I}^p(\boldsymbol{k})|. \tag{1}$$

Note that this normalization is defined by the library of images and does not change during the reconstruction process. We may define the normalized complex value of $k$-space intensities $y$ as,

$$y(\boldsymbol{k}) = \frac{\tilde{I}(\boldsymbol{k})}{\langle \tilde{I}(\boldsymbol{k}) \rangle_{n_1}}. \tag{2}$$

With this normalization, we may use the library of images to compute statistical properties that are useful for reconstruction. The average complex value of $y$ at any given location is simply given by $\mu(\mathbf{k}) = \mu'(\mathbf{k}) + i\,\mu''(\mathbf{k})$ where

$$\mu'_0(\mathbf{k}) = \frac{\sum_{p=1}^{n_1} Re[y(\mathbf{k})]}{n_1} \tag{3}$$

and

$$\mu''_0(\mathbf{k}) = \frac{\sum_{p=1}^{n_1} Im[y(\mathbf{k})]}{n_1}. \tag{4}$$

Similarly, we may define two covariances, namely $K'$ to reflect covariances between real terms and $K''$ to reflect covariances between imaginary terms. These covariances are defined between one location $\mathbf{k}$ and another location $\mathbf{k}'$ and can be computed using,

$$K'(\mathbf{k}, \mathbf{k}') = \frac{\sum_{p=1}^{n_1}\left(Re[y^p(\mathbf{k})] - \mu'_0(\mathbf{k})\right)\left(Re[y^p(\mathbf{k}')] - \mu'_0(\mathbf{k}')\right)}{n_1 - 1} \tag{5}$$

and

$$K''(\mathbf{k}, \mathbf{k}') = \frac{\sum_{p=1}^{n_1}\left(Im[y^p(\mathbf{k})] - \mu''_0(\mathbf{k})\right)\left(Im[y^p(\mathbf{k}')] - \mu''_0(\mathbf{k}')\right)}{n_1 - 1}. \tag{6}$$

Since each source $k$-space dataset was $N \times N$ pixels, we may consider $\mu_0'$ and $\mu_0''$ to each be vectors that are $N^2 \times 1$ while $K'$ and $K''$ are matrices that are each $N^2 \times N^2$. These dimensionalities are independent of the number of images in the training set.

We defined a series of kernel functions that multiplicatively modify $K'$ and $K''$. Specially, we propose kernel functions of the form $F(\boldsymbol{k}, \boldsymbol{k}')$ to define

$$G'(\boldsymbol{k}, \boldsymbol{k}') = K'(\boldsymbol{k}, \boldsymbol{k}') \cdot F(\boldsymbol{k}, \boldsymbol{k}'), \tag{7}$$

and

$$G''(\boldsymbol{k}, \boldsymbol{k}') = K''(\boldsymbol{k}, \boldsymbol{k}') \cdot F(\boldsymbol{k}, \boldsymbol{k}'). \tag{8}$$

Notably, the only statistical requirements for $F(\boldsymbol{k}, \boldsymbol{k}')$ to be valid are that it is positive semi-definite and $F(\boldsymbol{k}, \boldsymbol{k}') = F(\boldsymbol{k}', \boldsymbol{k})$. Specifically, we designed four kernel functions: (1) a unity function $F_u$, (2) a delta function $F_\delta$, (3) single envelope function $F_1$, and (4) double envelope function $F_2$. These functions can be expressed as:

$$F_u(\boldsymbol{k}, \boldsymbol{k}') = 1, \tag{9}$$

$$F_\delta(\boldsymbol{k}, \boldsymbol{k}') = \begin{cases} 1 & |\boldsymbol{k} - \boldsymbol{k}'| = 0 \\ 0 & |\boldsymbol{k} - \boldsymbol{k}'| \neq 0 \end{cases}, \tag{10}$$

$$F_1(\boldsymbol{k}, \boldsymbol{k}') = exp\left(-\frac{|\boldsymbol{k} - \boldsymbol{k}'|^2}{L_1^2}\right), \tag{11}$$

and

$$F_2(\boldsymbol{k},\boldsymbol{k}') = \frac{exp\left(-\frac{|\boldsymbol{k}-\boldsymbol{k}'|^2}{L_2^2}\right)+exp\left(-\frac{|-\boldsymbol{k}-\boldsymbol{k}'|^2}{L_2^2}\right)}{1+exp\left(-\frac{|\boldsymbol{k}-\boldsymbol{k}'|^2}{L_2^2}\right)\cdot exp\left(-\frac{|-\boldsymbol{k}-\boldsymbol{k}'|^2}{L_2^2}\right)}. \tag{12}$$

Here, $L_1$ and $L_2$ are hyperparameters that should be tuned based upon cross validation.

*B. Determination of Optimized Sampling Path Using Bayesian Optimization*

The goal of finding an optimum sampling path is to reduce the total uncertainty in our prediction of $\tilde{I}$. This was chosen as a goal because this uncertainty will directly translate to uncertainty in the final real-space image. Considering a case where $m$ points have been sampled, the goal becomes to predict the uncertainty at a new point to quantify how useful it would be to sample there. In particular, we consider that we have already sampled at points $k_{1:m}$ and these measurements have returned values $y_{1:m}$. Using Baye's rule, we can compute the posterior distribution of $k$-space intensity at an unobserved pixel at $\boldsymbol{k}$ as [25, 34]

$$y(\boldsymbol{k})|y_{1:m} \sim \text{Normal}\big(\mu'(\boldsymbol{k}), \sigma'(\boldsymbol{k})\big) + i \times \text{Normal}\big(\mu''(\boldsymbol{k}), \sigma''(\boldsymbol{k})\big) \tag{13}$$

with

$$\mu'(\boldsymbol{k}) = G'(\boldsymbol{k}, \boldsymbol{k}_{1:m})G'^{-1}(\boldsymbol{k}_{1:m}, \boldsymbol{k}_{1:m})\big(Re[y_{1:m}] - \mu'_0(\boldsymbol{k}_{1:m})\big) + \mu'_0(\boldsymbol{k}_{1:m}) \tag{14}$$

$$\mu''(\boldsymbol{k}) = G''(\boldsymbol{k}, \boldsymbol{k}_{1:m})G''^{-1}(\boldsymbol{k}_{1:m}, \boldsymbol{k}_{1:m})\big(Im[y_{1:m}] - \mu''_0(\boldsymbol{k}_{1:m})\big) + \mu''_0(\boldsymbol{k}_{1:m}), \tag{15}$$

$$\sigma'^2(\boldsymbol{k}) = G'(\boldsymbol{k}, \boldsymbol{k}) - G'(\boldsymbol{k}, \boldsymbol{k}_{1:m})G'(\boldsymbol{k}_{1:m}, \boldsymbol{k}_{1:m})^{-1}G'(\boldsymbol{k}_{1:m}, \boldsymbol{k}), \tag{16}$$

and

$$\sigma''^2(\boldsymbol{k}) = G''(\boldsymbol{k},\boldsymbol{k}) - G''(\boldsymbol{k},\boldsymbol{k}_{1:m})G''(\boldsymbol{k}_{1:m},\boldsymbol{k}_{1:m})^{-1}G''(\boldsymbol{k}_{1:m},\boldsymbol{k}). \tag{17}$$

In order to combine the real and imaginary terms into a single scalar uncertainty and convert these back into the unnormalized intensity, we calculate the intensity uncertainty $\sigma_I(\boldsymbol{k})$ as,

$$\sigma_I(\boldsymbol{k}) = \langle \tilde{I}(\boldsymbol{k}) \rangle_{n_1} \frac{\sqrt{\mu'^2(\boldsymbol{k}) \cdot \sigma'^2(\boldsymbol{k}) + \mu''^2(\boldsymbol{k}) \cdot \sigma''^2(\boldsymbol{k})}}{\sqrt{\mu'^2(\boldsymbol{k}) + \mu''^2(\boldsymbol{k})}}. \tag{18}$$

Importantly, in a Bayesian optimization context, $\sigma_I$ represents an acquisition function that prioritizes finding points with high uncertainty in $k$-space. As our priority is to obtain high fidelity images in real space, the use of this acquisition policy leverages the connection between point-wise uncertainty in $k$-space and image-wide uncertainty in real space as defined by the inverse Fourier transform as a linear operator.

While Equation (18) is useful for determining the uncertainty associated with a single point in $k$-space, our aim is to produce sampling paths that are compatible with conventional MRI systems that cannot necessarily choose sampling paths arbitrarily. Consistent with prior subsampling approaches, we instead consider sampling in circles with radius $r$ in $k$-space as these are feasible to realize in practice. In order to choose a circle to sample, we evaluated the angularly-averaged error $\langle \sigma_I \rangle_r$ on each possible circle, and select the previously unsampled circle with the highest $\langle \sigma_I \rangle_r$ as the next ring to sample. Subsequent rings can be selected until the allocated budget in terms of sampling density has been reached.

While the process for selecting an optimized path for a given image is deterministic, this relies on obtaining feedback from the image after selecting each subsequent ring. Such active selection during imaging may not be practical on all systems, thus we seek to identify generalized predetermined sampling paths that produce consistently high performance without feedback. To do this, we repeat the process of finding the optimized sampling path for all $n_2$ images in training subset 2. Once completed, the likelihood of a given $r$ value appearing in a given sampling path was computed for the whole subset. The generalized sampling path was then selected as the most commonly selected rings that together do not exceed the desired fraction of $k$-space. This subsampling mask is denoted $M(\mathbf{k}) \in [0,1]$.

*C. Reconstruction of Subsampled MR Image Using Bayesian inference*

The process of reconstructing a subsampled image began by selecting an image from the $n_3$ images in subset 3. The initial ground truth real-space image $I(\mathbf{x})$ was converted into $k$-space using an FFT to find $\tilde{I}(\mathbf{k})$. In order to approximate only measuring 12.5% of the $k$-space data, this complete $k$-space data was subsampled by multiplying it by the binary matrix $M(\mathbf{k})$ corresponding to the generalized sampling path. Importantly, this process did not require any feedback as the complete sampling path was known ahead of time, as learned based on training sets $n_1$ and $n_2$. This subsampled data was then normalized according to Equation (1) to find $y(\mathbf{k})$. Subsequently, reconstruction proceeds by using Equations (8) and (9) to find $\mu'(\mathbf{k})$ and $\mu''(\mathbf{k})$ using only the $m$ non-zero terms of $M$. The final reconstructed image therefore,

$$\tilde{I}^*(\mathbf{k}) = \left[\left(1 - M(\mathbf{k})\right) \times \left(\mu'(\mathbf{k}) + i\mu''(\mathbf{k})\right) + M \times y(\mathbf{k})\right] \times \langle I(\mathbf{k}) \rangle_{n_1}, \tag{19}$$

which uses the sampled values where known and fills in inferred values in all other locations in $k$-space. Finally, we use an IFFT of $\tilde{I}^*(\boldsymbol{k})$ to compute the reconstructed real-space image $I^*(\boldsymbol{x})$, which is $N \times N$ pixels.

D. Experimental Setup and Procedure

1) MR Image Data

In order to explore the ability of GP-based sampling and reconstruction to effectively subsample MR images, we first needed to establish a training set of data. In particular, we selected the IXI dataset, which contains 600 3D MRI datasets of healthy brains, some taken at 1.5 T and some taken at 3.0 T [36]. To partition the dataset into testing and training portions, we selected 7,391 2D $256 \times 150$ pixel axial images $I(\boldsymbol{x})$ from T1-weighted 3D MR datasets in NIFTI format, reshaped them into $N_0 \times N_1$ pixel images, where $N_0 = 256$ and $N_1 = 200$, and then zero-padded the left and right edges to make the images into $N_0 \times N_0$ squares with each pixel representing 1.2 $\times$ 1.2 mm$^2$. For each image, the intensity $I(\boldsymbol{x})$ was normalized by its $\max(I(\boldsymbol{x}))$. These images were then converted to $k$-space data $\tilde{I}(\boldsymbol{k})$ using a FFT and cropped to retain the central $N \times N$ ($N = 160$) region for computational efficiency. These $k$-space data were randomly divided into $n_1 = 6{,}822$, $n_2 = 69$, and $n_3 = 500$ partitions for training and testing. Additional diffusion weighted images were collected as 2D MR images using a Philips Achieva 1.5 T system with a gradient $b = 1000$ s/mm$^2$ (b1000) and the apparent diffusion coefficient map (ADC) was calculated.

2) Setup and Evaluation

The final reconstructed $k$-space data $\tilde{I}^*(\boldsymbol{k})$ was zero-padded into $N_0 \times N_0$ pixels and then IFFT was used to compute the reconstructed real-space image $I^*(\boldsymbol{x})$. Image reconstruction quality was evaluated using normalized mean square error (NMSE), a measure of the mean greyscale value difference between the target and reconstructed image, and structural similarity index metric (SSIM), which measures the similarity between the image pair by calculating the inter-dependency among nearby pixels using a sliding window [37]. Code to compute FFT, IFFT, and reconstruction metrics were taken from the Facebook research fastMRI project [37]. Numerical experiments are carried out in a Tesla V100 GPU with 16 GB RAM.

*3) Experimental Procedure*

We calculated $\mu_0'(\mathbf{k})$, $\mu_0''(\mathbf{k})$, $G'(\mathbf{k},\mathbf{k}')$, and $G''(\mathbf{k},\mathbf{k}')$ using training set 1 and stored these for subsequent BO. Following the methods in part A-C, training set 2 was used to find the optimal subsampling pattern with less than 12.5% (8-fold) $k$-space information for different $G'(\mathbf{k},\mathbf{k}')$ and $G''(\mathbf{k},\mathbf{k}')$ using those four kernel functions in Equations (9)-(12). Hyperparameters $L_1$ and $L_2$ in Equation (11) and Equation (12) were tuned based on the minimum average NMSE value of 25 randomly drawn images from training set 2. Using the different optimized sampling paths and kernel functions, we reconstructed $k$-space data based on Equation (13) to evaluate the reconstruction quality of the testing set. Additionally, in order to look at how the training data size affects the hyperparameters and image reconstruction process, we reduced $n_1$:$n_2$ from 6822:69 to 1722:17 and ultimately 431:5 and repeated the same process described above. Finally, we used the trained model based upon IXI healthy brain MR images in T1 mode and tested the reconstruction quality of example pathological brain images with strokes.

## 3. Results

*A. Subsampling Paths with Different Covariances*

In order to begin to evaluate the use of GP-based reconstruction, it is first necessary to identify optimized sampling paths. To do this, we first use training set 1 to define the library of means and covariances by computing $\mu_0'(\boldsymbol{k})$, $\mu_0''(\boldsymbol{k})$, $G'(\boldsymbol{k}, \boldsymbol{k}')$, and $G''(\boldsymbol{k}, \boldsymbol{k}')$. Importantly, these terms can be used together with Eq. (18) to compute $\sigma_I$, or the expected uncertainty in $\tilde{I}(k)$. Not only can this term be computed prior to taking any measurements, but it can also be computed after performing measurements at any array of points. In particular, considering a subsampling path that consists of a set of $j$ circles, this subsampling path can be used to predict the predicted uncertainty $(\sigma_I)_j$ at any subsequent potential location to sample, as shown in Figure 2a.

Having formalized the process for updating the uncertainty, optimized sampling paths can be chosen by sequentially selecting the ring with radius $r$ with the largest $\langle(\sigma_I)_j\rangle_r$. Since these values are peaked near $r = 0$, this process sequentially moves to larger values of $r$, as shown in Figure 2b. While this process is stochastic, we repeated it for each image in training set 2 to produce a probability each $r$ was selected (Figure 2c). These data in hand, a general sampling path may be selected as the most probable radii that together cover less than 12.5% of $k$-space.

While the process described in Figure 2 shows the optimal path determination process found by directly using all data in the empirical covariance matrices and means, we hypothesized that this introduces a large amount of noise due to the limited size of the dataset. To illustrate this, Figure 3a shows a section of $|K'(\boldsymbol{k}, \boldsymbol{k}')|$ vs. $\boldsymbol{k}'$ with $\boldsymbol{k} = (40, -40)$. Notably, there exist two peaks that denote regions with strong correlations, the first is around $\boldsymbol{k} = \boldsymbol{k}'$, which we term the direct point, and the other is near $\boldsymbol{k}' = \boldsymbol{k}^* = (-40, 40)$, which is its Hermitian point. Away from these points, the $K'$ is much smaller in magnitude, although a line cut between these points (Figure 3b) reveals appreciable noise in regions away from the peaks. We posit that these fluctuations represent noise that should not be included in the reconstruction process.

To explore the hypothesis that regions of the covariance matrix should be removed to facilitate reconstruction, we define a series of kernel functions that modify the covariance matrix to different degrees. For instance $F_\delta$ is defined by a delta function that rejects all data besides $\boldsymbol{k} = \boldsymbol{k}'$, which effectively amounts to not using any covariance information. Conversely, we also define $F_u$, which is unity, or preserves the full covariance matrix as in Figure 2. However, to preserve the physical significance of the covariance matrix, we also introduce $F_1$, which keeps a region defined by a Gaussian around $\boldsymbol{k} = \boldsymbol{k}'$. Further, we introduce a kernel function $F_2$, which also includes a Gaussian around the Hermitian point. Importantly, these latter two envelope functions include hyperparameters, namely the widths of their Gaussian peaks. These were chosen based on the minimum average NMSE value of 25 randomly drawn images from training set 1. The final tuned values are $L_1 = 15$ and $L_2 = 13$.

In order to understand whether the kernel functions would affect the generalized sampling paths, the process described in Figure 2 was repeated using either $F_1$ or $F_2$. Interestingly, while the sampling path found using $F_1$ identified a central circle as being important to sample, the radii did not extend beyond $r = 66$. We attribute this focus on the central region as owing to the omission of some important information, namely from the Hermitian point (Fig. 3d). In contrast, the generalized sampling path predicted using $F_2$ stretched nearly to $r = 80$, indicating that the algorithm was confidently learning most of the image (Fig. 3e). Based on these results, $F_2$ appeared to be the most promising kernel to guide subsampling and reconstruction.

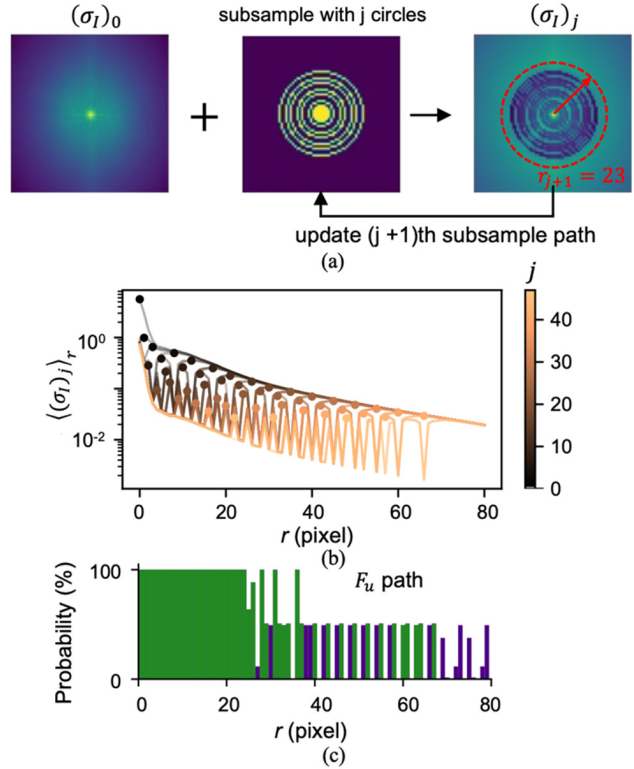

**Fig. 2. (a) Schematic showing the process of choosing the optimal circles to sample using BO. Each step involves combining the initial error in k-space magnitude $(\sigma_I)_0$ with a subsample of $j$ circles. These are used with the multivariate normal distribution to infer a post-measurement error $(\sigma_I)_j$. Finally, the unsampled circle with $k$-space radius $r$ with maximum angularly averaged error $\langle\sigma_I\rangle_r$ is selected as the next circle to add to the sampling path, shown here in red. (b) $\langle(\sigma_I)_j\rangle_r$ vs. $r$ shown at 48 different values of $j$. Here, the dots indicate the largest value that was chosen for the $j+1$ circle. (c) histogram of sampled radii for all all $n_2 = 69$ images showing the frequency of appearance. Green bars were the most common circles that together formed 12.5% of $k$-space. They were selected as the optimized circular sampling path.**

## B. Reconstruction Results with Different Covariances

In order to evaluate the quality of MR images subsampled and reconstructed using the four kernel functions $F_u, F_\delta, F_1$, and $F_2$, we examined a single image subsampled to 12.5% and reconstructed using these approaches. As shown in Figure 4a, each column shows the general subsampling path, complete reconstructed image, and a magnified region of interest. The quality of the reconstructed images, as exemplified by the fine details in the magnified region, improves as more advanced kernel functions are used with $F_2$ exhibiting the best performance.

As a more objective measure of reconstruction quality, maps of the normalized error show a dramatic decrease by a factor of ~2 in going from the $F_\delta$ kernel to the $F_2$ kernel. This improvement can be rationalized in that the $F_2$ method obtains the best balance of accepting significant correlations while rejecting noise. To determine the degree to which this improvement was consistent across the testing set, the average NMSE was computed for each of the 500 reconstructed images in the testing set (Fig. 4b,c). As with the individual image, we find that $F_\delta$ exhibits the highest noise at any image sampling percentage larger than 5%. Interestingly, at low sampling percentages, $F_1$ exhibits worse performance than $F_u$, showing that in conditions with little data, it is better to accept more noise in favor of retaining the Hermition point. As expected, $F_2$ exhibited the best performance at all sampling ratios, decreasing to an NMSE of 0.00252 at 12.5% sampling. Interestingly, the computed average SSIM reveals a similar trend with $F_2$ leading to the best performance and a peak SSIM of 0.963 at 12.5% sampling. The other methods revealed some interesting compromises where $F_u$ became the worse performer at high sampling percentage, showing the peril of accepting too much noise. Importantly, the best NMSE and SSIM metrics obtained for $F_2$ at 12.5% reconstruction are superlative among a survey of other methods (Table I), including those that incorporate DL. [19, 37-40]

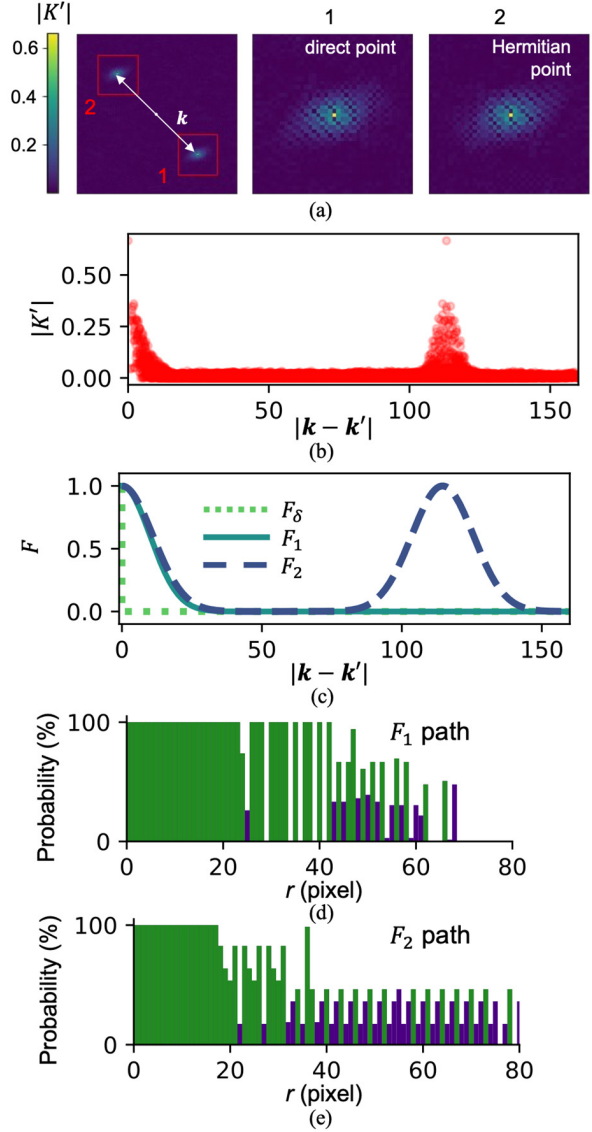

**Fig. 3.** (a) Image showing a map of multivariate normal distribution covariance $|K(k, k')|$ vs. $k'$ with $k = (40, -40)$. The image shows two peaks, one at $k' = k$ (direct point), and the second at $k' = k^*$ (Hermitian point), which represents the Hermitian symmetry of $K$. Insets show zoomed-in images of the two peaks. (b) Plot showing a profile of $|K|$ vs. $|k - k'|$ for $k = (40, -40)$ showing the direct and Hermitian points as peaks. (c) Proposed kernel functions $F_\delta$, $F_1$, and $F_2$ vs. $|k - k'|$. $F_1$ only depends on $|\vec{k}_1 - \vec{k}_2|$ while $F_2$ is shown for $\vec{k}_2 = (40, -40)$. (d-e) Outcome of BO to find optimal paths with histograms representing the most commonly sampled radii when studying $n_2 = 69$ test images. Green bars were the most common circles that together formed 12.5% of k-space. Panel d was computed using $G_1$ as covariance. Panel e was computed using $G_2$ as covariance.

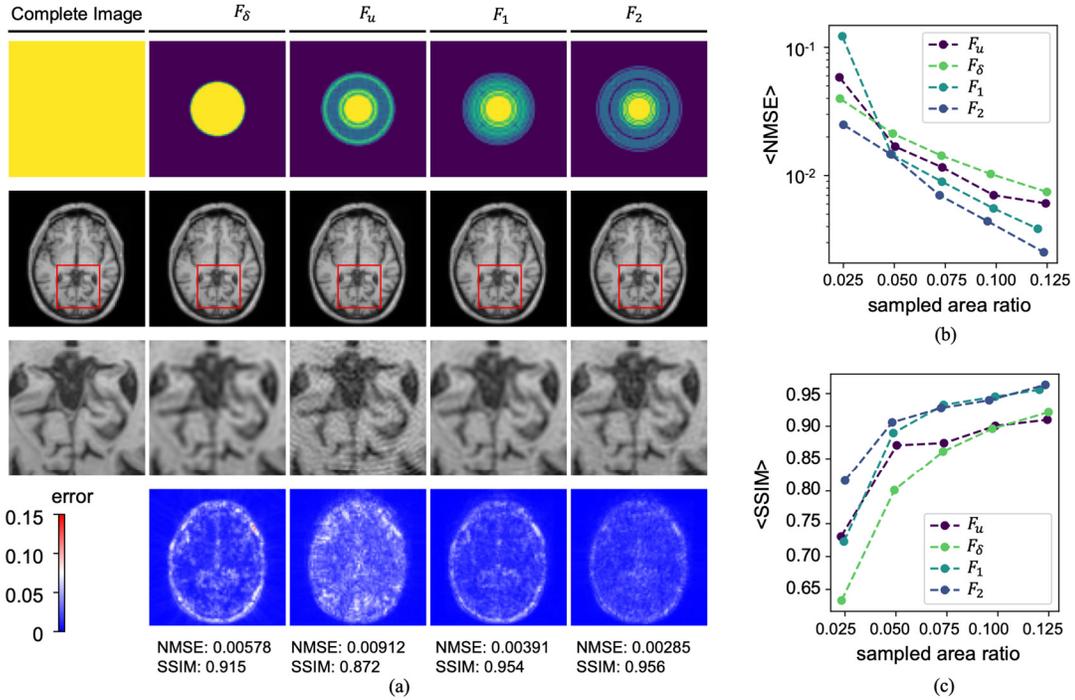

**Fig. 4.** (a) Comparison of reconstruction quality of a single image reconstructed from 12.5% of the k-space image. Four envelope functions are compared, each trained with $n_1 = 6822$ and $n_2 = 69$. The first column shows the ground truth image and magnified inset to show detail. Subsequent columns show the reconstructed image produced by each of the four methods. The bottom row shows a map of absolute error for the zoomed in-region. Normalized mean square error (NMSE) and structural similarity index measure (SSIM) are reported for each of the reconstructed images. (b,c) Average NMSE and SSIM shown for each of the four reconstruction methods vs. the total area in k-space used for reconstruction, respectively.

|  | Acceleration | Dataset | NMSE | SSIM |
|---|---|---|---|---|
| Multicoi-Unet | 8X | fastMRI | 0.0443 | 0.884 |
| AIRS Medical | 8X | fastMRI | 0.0035 | 0.951 |
| MDNNSM | 8X | fastMRI | 0.005 | 0.944 |
| Self-Supervised | 8X | IXI | NaN | 0.927 |
| Deep Convolutional Encoder-Decoder | 5X | IXI | 0.0793 | 0.942 |
| GP DE (This Paper) | 8X | IXI | 0.00252 | 0.963 |

**Table I.** Comparison of average NMSE and SSIM using different methods at an 8-fold $k$-space acceleration factor.[19, 37-40]

## C. Reconstruction with Different Training Dataset Sizes

While the GP-based subsampling and reconstruction process was found to be very successful, we sought to explore the degree to which this was based upon the volume of data available. In particular, we hypothesized that additional data allows the empirical covariance functions to be learned with higher precision, thus increasing reconstruction quality. In order to explore this, we repeated the training, subsampling, and reconstruction process two more times with datasets that were truncated by roughly 4 and 16 fold. In particular, we studied the cases where $n_1 = 1722$ and where $n_1 = 431$ using the $F_1$ envelope function. Interestingly, when we repeated this process, we first found that the hyperparameter $L_1$, which defines the width of the kernel function, varied with the dataset size. This can be directly seen by visualizing a region of the covariance function centered at $|\mathbf{k} - \mathbf{k}'| = 0$ for the three dataset sizes (Fig. 5a), which reveals a more gradual fall-off away from the direct point for the largest data set. As a result, we found that $L_1$, while at 10 for the largest data set, decreased to 9 and 6 for the medium and small datasets, respectively. These values confirm the hypothesis that the volume of data in the training library determines the degree the extent of the region in $k$-space that can be inferred from each measurement.

In order to determine the consequences of the varying dataset size in terms of reconstruction performance, we repeated the sample reconstruction process using the two additional dataset sizes. As shown in Fig. 5b, we found that while $\langle NMSE \rangle$ did not appreciably change, and $\langle SSIM \rangle$ increased slightly with increasing dataset size. Besides reconstruction quality based on these metrics, we noted a striking increase in image reconstruction computational time with decreasing $n_1$ (Fig. 5b). We rationalize this increase as smaller values of $L_1$ value together with larger noise leading to more near-zero off-diagonal terms in $\widetilde{K}'(\mathbf{k}_{1:M}, \mathbf{k}'_{1:M})$ and $\widetilde{K}''(\boldsymbol{k}_{1:M}, \boldsymbol{k}'_{1:M})$ that increased the data instability and burdened the matrix inversion process in Eq. (14)-(15). This trend where more training data leads to faster reconstruction is in some ways counterintuitive as typically more data corresponds to higher model complexity and therefore longer evaluation times. Here, however, the model is fixed and the additional data simply serves to improve estimates of variances and covariances, which accelerates and improves reconstruction.

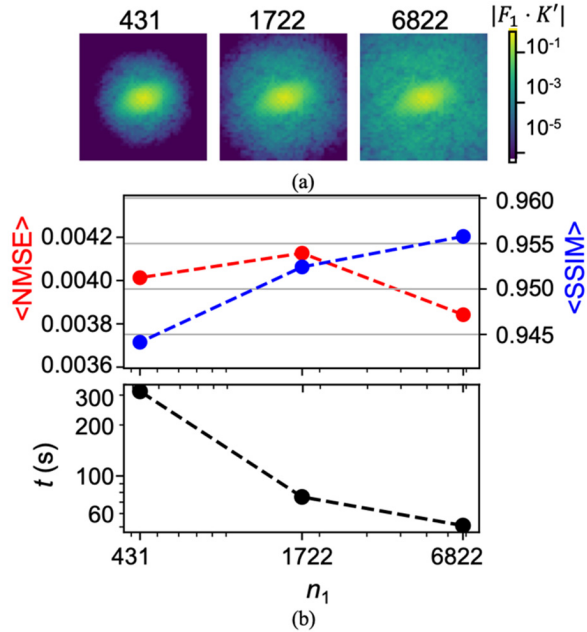

Fig. 5. (a) Image showing zoomed-in slices of $|F_1 \cdot K'(k, k')|$ vs. $k'$ with $k = (40, -40)$. Images are centered on $k = (40, -40)$ and show a 60×60 pixel region. Each image is produced using a different $n_1$ and the optimum envelope width is chosen for each. (b) Average NMSE and SSIM vs. $n_1$ for reconstructions performed using the $F_1$ envelope. Average time $t$ to compute perform reconstruction vs. $n_1$ for reconstructions performed using the $F_1$ envelope.

D. *Reconstruction of Pathological MR Images*

A key question about the application of reconstruction methods is whether they can function for data types outside the scope of the training data. For example, despite the present GP-based model being trained exclusively on images of healthy brains taken using a T1-weighted
22

sequence, we sought to determine whether this model could provide useful guidance for images of pathological brains taken using other means. In particular, we used the subsampling path and reconstruction algorithms described in Section III.B to study an example b1000 and ADC image from diffusion weighted imaging of a brain with a recent stroke. Like the images of healthy brains, the NMSE was minimized and SSIM maximized when using the $F_2$ kernel for reconstruction (Figure 6a).

While image quality metrics are important, perhaps more important are whether indicators of disease state are readily visible in the reconstructed image. To test this, we visualized the stroke region as shown by both b1000 and ADC images (Figure 6c). Here, the stroke is visible as the region that is bright on the b1000 images while being dark on the ADC images. Strikingly, not only was this domain readily apparent in the image reconstructed from 12.5% of the image using $F_2$ kernel, but it was even apparent when using only 2.5% of the image for reconstruction. This result shows two important findings, (1) that a model trained on a different imaging sequence and field strength can provide value and (2) that this model is useful for exploring pathologies beyond the training set, here shown in the context of stroke segmentation.

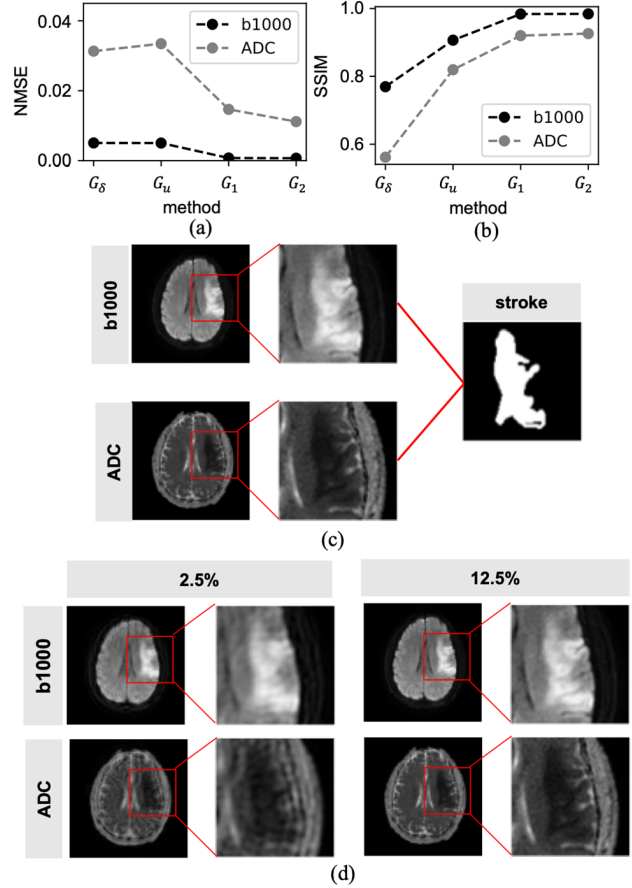

**Fig 6. (a,b) NMSE and SSIM shown for each of the four reconstruction methods for an example 2D brain slice in two different MRI scanning modes: "b1000" and "ADC". (c) "b1000" and "ADC" ground truth images with magnified inset to show the stroke area. Those two images combined help define the final stroke map, shown on the right panel. (d) Reconstructed images with double envelope methods ($F_2 \cdot K$) with 2.5% or 12.5% sampled area.**

## 4. Discussion

In this work, we proposed a method for accelerating MRI using a Bayesian framework to both select optimal *k*-space subsampling paths and reconstruct the resulting subsampled images. In contrast with conventional applications of Bayesian optimization which iteratively construct the covariance matrix by tuning kernel functions to learn the correlation between data points, we employed an empirical mean and covariance function based on a set of training images to serve as a reference library for Bayesian inference. A unique feature of our implementation of BO is that while conventionally this must be performed as a closed-loop process in which data is used to iteratively select new points to sample, here we determine optimal generalized sampling paths that allow a fixed subsampling pattern to be used in an open-loop setting. This is a crucial innovation in that it means that special software to interact with an MRI system is not needed. Moreover, the use of concentric rings as the sampling paths is crucial as it facilitates adoption on existing commercial systems. Compared with other methods for accelerating MR imaging using DL active acquisition methods which require complex subsampling paths that are determined in a closed-loop setting [22-24], this subsampling pattern is easily realizable due to prior theoretical and experimental studies on the MR pulse sequences needed for circular sampling [7, 9, 41].

A crucial innovation of this work is that the unsampled $k$-space area can be populated using Bayesian inference. While directly using the empirical covariance for reconstruction provided limited benefit, we employed a variety of non-stationary kernel functions that balanced accepting valid correlations and rejecting undesirable noise from the training data. Best among these kernel functions was a double envelope function in which points close to the sampled point and those with Hermitian symmetry were accepted. Importantly, we found that the more data that was provided in the training set meant that longer ranged correlations could be leveraged, providing a strong incentive to build large datasets for training. Along these lines, it is important to emphasize that once training is complete, models trained with more data are not any larger and in fact can be used to reconstruct images faster. There is also only a single hyperparameter to tune (the length scale of the kernel function), which facilitates rapid training. This implementation of Bayesian optimization opens the door to other methods taken from this discipline to accelerate the learning process. For instance, other acquisition policies can be employed, such as those that examine differences in entropy rather than uncertainty. Further, the Gaussian envelope functions used here could be replaced with other functions such as exponential or Matérn functions that carry fewer assumptions about the differentiability of the underlying data.

Perhaps the most important outcome of this work is superlative reconstruction accuracy. In particular, we found that it was possible to obtain $<NMSE> = 0.00252$, and $<SSIM> = 0.963$ for the 500 T1 mode 2D MR images of healthy brains while only sampling 12.5% of the image. Not only was this method found to be powerful for reconstructing T1-weighted images of healthy brains, but the method trained on these images could be extended to images of brains with strokes collected using a different pulse sequence to provide clinically relevant information with only 2.5% subsampling. This transferability is a powerful motivation to continue to explore this approach.

Looking ahead, we believe that the incorporation of additional datasets, imaging modalities, and pathologies could enhance this method further. Given the growing importance of obtaining rapid information and the emerging field of low-field MRI in which low signal-to-noise ratios necessitate longer collection times, this subsampling and reconstruction process could find wide application across medical imaging.

**Acknowledgements**

This work was supported by the Rajen Kilachand Fund for Integrated Life Science and Engineering and the authors acknowledge Dr. Marcus Noack for helpful discussions.

**Conflict of Interest Statement**

The authors declare no conflict of interest.